# Indistinguishable photons from the resonance fluorescence of a single quantum dot in a microcavity

Serkan Ates[1], Sven Marcus Ulrich[1], Stephan Reitzenstein[2], Andreas Löffler[2],

Alfred Forchel[2] and Peter Michler[*1]

[1] *Institut für Halbleiteroptik und Funktionelle Grenzflächen, Universität Stuttgart, Allmandring 3, 70569 Stuttgart, Germany.*

[2] *Technische Physik, Universität Würzburg, Am Hubland, 97074 Würzburg, Germany.*

*Email: p.michler@ihfg.uni-stuttgart.de; Homepage: http://www.ihfg.uni-stuttgart.de

**Novel quantum information processing schemes like linear optics quantum computation and quantum teleportation are based on the effect of two-photon interference of two single-photon pulses[1-5] on a beamsplitter. An essential prerequisite for the successful realization of these applications is the possibility to generate *indistinguishable* photons. For pulsed operation, the two photons have to be *Fourier transform-limited* and identical in terms of pulse width, spectral bandwidth, carrier frequency, polarization, transverse mode profile, and arrival time at the beamsplitter. The critical ingredient to create such 'ideal' photons is the process of initial excitation of the emitter, which strongly influences the coherence properties and, consequently, the indistinguishability of the emerging photons. So far, indistinguishability tests by two-photon interference studies have been reported from different model systems like single atoms[6], trapped ions[7], molecules[8], and semiconductor quantum dots[9], all of which were subject to an**

**incoherent** *preparation of their corresponding radiative states. Here we demonstrate purely resonant continuous-wave optical laser excitation to coherently prepare an excitonic state of a single semiconductor quantum dot (QDs) inside a high quality pillar microcavity. As a direct proof of QD resonance fluorescence, the evolution from a single emission line to the characteristic Mollow triplet[10] is observed under increasing pump power. By controlled utilization of weak coupling between the emitter and the fundamental cavity mode through Purcell-enhancement of the radiative decay, a strong suppression of 'pure' dephasing is achieved, which reflects in close to Fourier transform-limited and highly indistinguishable photons with a visibility contrast of 90%. Our experiments reveal the model-like character of the coupled QD-microcavity system as a promising source for the generation of 'ideal' photons at the quantum limit. From a technological perspective, the vertical cavity symmetry -- with optional dynamic tunability[11] -- provides strongly directed light emission which appears very desirable for future integrated emitter devices.*

If the radiative transition between two quantum states of a single emitter is broadened solely by the spontaneous radiative emission process, the single photon pulses are Fourier transform-limited, i.e. the characteristic temporal widths of the photons are $T_2 = 2/\gamma_0$ with a natural linewidth of $\gamma_0$. In the presence of dephasing processes though, the photon is in an incoherent mixed state as opposed to a pure state. Here $T_2$ can be considered as the dephasing time of the excited state and is defined by $1/T_2 = 1/(2T_1) + 1/T_2^*$, where $T_1$ is the radiative lifetime of the emitter, and $T_2^*$ is the pure dephasing time (i.e. loss of coherence without recombination). Only in the Fourier-limited case one obtains $T_2/2T_1 = 1$, i.e. coherence being purely defined by the radiative decay.

Incoherent pumping of a solid-state single quantum emitter typically leads to homogeneous broadening of the excited state and therefore a reduction in coherence. Additionally, non-radiative relaxation from the excited state to the emitting ground state introduces a time jitter in the photon emission process. Both effects reduce the two-photon interference and have been theoretically analyzed[12]. Therefore, a true resonant excitation process appears crucial for the generation of *indistinguishable* photons. The challenge for resonant pumping in solid state is the separation of the strong laser excitation light from resonantly scattered photons. Coherent excitation of a molecule emitting transform-limited photons has been recently realized[13]. However, background laser scattering was of the same order of magnitude as the molecular fluorescence. True resonant excitation (s-shell) of a single quantum dot in a planar cavity has also been demonstrated recently[14]. In both studies, two-photon interference measurements have not been performed.

The single-photon generation scheme that we use is based on individual self-assembled (In,Ga)As/GaAs QDs embedded in a high-quality microcavity structure. To ensure Fourier transform-limited photon emission with negligible background we use an orthogonal excitation and detection technique and stabilized low-temperature operation at $T \geq 10K$ (Fig. 1a). We also benefit from the Purcell effect which leads to a reduction of $T_1$, thus reducing the impact of possible phonon dephasing processes[15]. Furthermore, the efficient coupling of the QD emission into the cavity mode and the enhanced photon collection effect out of the microcavity serves to increase the signal-to-noise ratio representing the key issue for performing two-photon interference measurements with high visibilities.

Individual microcavities are optically addressed close to the cleaved edges of our sample structure (details given in[16] and 'methods'). After spatial filtering of the collected µ-PL by a pin-hole and spectral signal filtering by a monochromator, our setup provides high-resolution PL (HRPL) measurements by a scanning Fabry-Pérot interferometer in combination with photon statistics analyses in terms of $g^{(2)}(\tau)$ 2nd-order correlations and two-photon interference experiments (see methods).

Fig. 2a shows a µ-PL spectrum of a 1.75 µm pillar cavity at T = 10 K under cw p-shell excitation. The PL peak at 1.357 eV is identified as the single exciton recombination (X) whose fine structure splitting becomes clearly visible in a Fabry-Pérot HRPL scan (Fig. 2b). The fine-structure splitting is $\Delta E_{FS} = 11.0 \pm 0.2$ µeV and the FWHM of the individual fine-structure components are 7.5 and 7.6 µeV, respectively. The fundamental mode (FM) of the cavity is mainly fed by non-resonantly coupled single excitonic emission[17-19] and some background contribution.

By temperature-tuning we are able to shift the X transition into resonance with the FM of the pillar cavity (see also supplementary information, Fig. S1). This allows us to systematically decrease the radiative lifetime $T_1$ through the Purcell effect, and also enhance the photon collection efficiency. The resonance condition is reached at 24 K. Fig. 2c shows the measured µ-PL decay times as a function of the QD-FM detuning $\Delta E$. For this measurement, the sample was quasi-resonantly excited (p-shell) with 2 ps pulses of a mode-locked Ti:sapphire laser (76 MHz). Symmetric in detuning around QD-mode resonance, a pronounced decrease of the decay time from ~ 820 ± 10 ps at $\Delta E$ = 250 µeV down to ~ 65 ± 10 ps could be traced, reflecting a large Purcell effect of factor ~ 12.

Fig. 2d shows the result of a cw-laser resonance scan (step width $\Delta f_L = 260$ MHz) over the X transition of the same single QD under synchronous detection of the emission. In this signal, which is composed of QD resonance fluorescence and scattered laser light with a high signal-to-noise ratio of factor ~ 5, we obtain a well resolved double peak structure. The doublet nicely reflects the fine-structure splitting $\Delta E_{FS}$ already measured in HRPL under p-shell excitation (Fig. 2b). Despite a slightly increased sample temperature of 18 K we now obtain reduced FWHM values of the two components of only 6.1 ± 0.2 and 3.7 ± 0.2 µeV, respectively. This already indicates a reduced dephasing with respect to p-shell excitation. In Fig. 2d the occurrence of both fine-structure components in emission suggests that the main axes of the studied QD are tilted with respect to the cleaved [100] edge of the sample. Otherwise, only one FS component would have been excited with the horizontally polarized laser under side excitation. For all subsequent measurements, pure linearly polarized resonant-excitation into the high-energy component of the X resonance has been applied. In addition, photon detection was limited to the corresponding linear polarization axis to further suppress the other fine structure component and laser stray light.

As was originally derived in the theoretical work by B. R. Mollow[10] on the interaction of atomic transitions with a monochromatic resonant driving (laser) light field, the conditions of strong resonant excitation 'dress' a radiative state into a quadruplet of two excited and two relaxed electronic states. Accordingly, four possible radiative transitions exist, two of which are spectrally degenerate at frequency $\omega_0$ (Fig. 3b). Therefore, the original photon emission channel (frequency $\omega_0 = \omega_{Laser}$) is characteristically decorated by spectral satellites at frequencies $\omega_0 \pm \Omega$ (with Rabi frequency $\Omega$). Recently, the observation of such Mollow triplet signatures has been achieved from high-resolution laser absorption spectroscopy[20,21] and resonance fluorescence[22] on single QD structures.

Fig. 3a shows an excitation power-dependent Fabry-Pérot HRPL scan series of resonance fluorescence from the single QD in Fig. 2a, observed at 10 K. In the regime of low excitation powers, a purely Lorentzian line shape with a constant width of $\Delta_{FWHM}$ = 1.15 ± 0.05 µeV is obtained in emission. We emphasize that under these conditions the linewidth is now found to be maximally reduced by a factor of ~ 7 as compared to the emission under p-shell excitation at the same temperature (10 K; Fig. 2b). In total, this demonstrates the pronounced reduction of dephasing processes achievable under resonant s-shell excitation and low temperatures. For this low excitation limit $\Omega^2 \ll 1/(T_1T_2)$ of the linewidth[14], a long coherence time of $T_2 = 1150 \pm 50$ ps can be derived.

Second-order correlation measurements under these resonant excitation conditions have revealed a pronounced antibunching dip with $g^{(2)}(0) = 0.19$ (Fig. 4a). Deconvolution with the time resolution of our HBT setup (~ 400 ps) gives a value of $g^{(2)}(0) = 0.08$, demonstrating the nearly pure single-photon nature of the collected signal (background ~ 4%) at this excitation power regime. Using a PL decay time of $630 \pm 20$ ps from our time-resolved measurements of Fig. 1c as $T_1$, we can infer a $T_2/2T_1$ ratio of $0.91 \pm 0.05$. We want to point out that the 'true' lifetime $T_1$ should be slightly shorter than the measured decay time, thus creating a $T_2/2T_1$ ratio even closer to the ideal value of 1. This is because the non-radiative relaxation time (~ 10–50 ps)[23] from the excited state (p-shell) to the emitting state (s-shell) slightly increases the experimentally measured effective decay time with respect to the true $T_1$ time. We therefore conclude, that close to ideally Fourier-transform limited single photons with > 90% fidelity have been generated and verified.

Under increasing excitation powers, symmetric satellite peaks with increasing energy separation appear in the resonant QD emission spectrum (Fig. 3a). A large Rabi splitting between these side and central peaks of $\hbar\Omega = 26.7$ µeV ($\Omega/2\pi = 6.4$ GHz) is obtained, which is about a factor of two larger than corresponding values recently reported by Atature and coworkers[22]. Plotting the side-to-central peak separation as a function of the effective Rabi frequency, i.e. proportional to the square root of excitation power, an almost perfect proportionality is found in accordance with the original theory by Mollow[10] (Fig. 3c).

It is important to note that the laser background contribution considerably increases with excitation power and finally dominates the central feature of the Mollow spectrum. This effect is due to the non-linear absorption behaviour of a resonantly excited two-level system (QD) which results in fluorescence emission saturation, while the laser signal increases linearly with excitation power. For our experiments we therefore chose a low excitation power (where $\hbar\Omega \sim 0.9$ µeV; according to ~ 80% of the saturation level) for our photon correlation and two-photon interference measurements which are discussed below. In addition, the signal-to-background ratio was improved by decreasing the QD-FM detuning and utilizing the increasing Purcell effect (see Fig. 1c) together with enhanced QD-mode coupling. Since we use temperature tuning, the $T_2$ time decreases at the same time due to an increased dephasing by phonon interaction. Therefore, we selected an intermediate QD-FM detuning of $\Delta E = 190$ µeV (with $T_1 = 560 \pm 20$ ps) where a linewidth measurement in the low intensity limit (Fig. 2d) yields a FWHM of 3.7 µeV (T = 18 K), corresponding to $T_2 = 360$ ps. With this we achieve a signal-to-background ratio of ~ 95 %. The chosen detuning leads to a reduction of $T_2/2T_1$ to 0.32. However, this does not influence the principally measurable visibility of the two-photon interference experiment discussed in the following since under cw excitation the maximum observable visibility is given by the ratio $T_2/(2 \delta t)$, where $\delta t$ is the response time of the detectors[24].

Two-photon interference measurements were carried out with the fiber-based Mach-Zehnder interferometer shown in Fig. 1a. Figs. 4b and c depict the results of the two-photon interference measurements $g^{(2)}_\perp(\tau)$ and $g^{(2)}_\parallel(\tau)$, which describe correlations between photons with orthogonal and parallel polarizations, respectively. In orthogonal configuration, the two paths are distinguishable and one expects $g^{(2)}_\perp(0) = 0.5$, whereas in the parallel configuration the paths are indistinguishable and $g^{(2)}_\parallel(\tau)$ should become zero in the ideal case[8,24]. We observe a value of $g^{(2)}_\perp(0) = 0.55 \pm 0.01$ and $g^{(2)}_\parallel(\tau) = 0.22$

± 0.01 for the orthogonal and parallel cases, respectively. In addition, two nearly equal correlation dips down to ~ 0.75 are observed at τ = ±13 ns due to the delay in our interferometer, indicating a balanced beamsplitter.

The two-photon interference visibility can be defined[8,24] as $V_{HOM}(\tau) = [g^{(2)}_{\perp}(\tau) - g^{(2)}_{\parallel}(\tau)] / g^{(2)}_{\perp}(\tau)$. Fig. 4d shows the visibility curve with a maximum normalized value of 0.6 (instrumental response-convoluted) at τ = 0. In the following we analyze and discuss the results of the photon correlation measurements. In a two-level coherent excitation the second-order coherence function $g^{(2)}(\tau)$ is determined by both $T_1$ and $T_2$ (in contrast to the incoherent situation where only $T_1$ determines the shape of the antibunching notch at low pump power[13,25-26]). Under strong coherent excitation, Rabi oscillations are expected to appear in the correlation function that persist within the coherence time, as was recently demonstrated by Muller et al.[27].

Explicitly, $g^{(2)}(\tau)$ is given by

$$g^{(2)}(\tau) = 1 + \frac{\lambda_-}{2q} e^{|\tau|\lambda_+} - \frac{\lambda_+}{2q} e^{|\tau|\lambda_-}$$

where

$$\lambda_{\pm} = -\frac{1}{2}(\gamma_0 + \gamma/2) \pm q \text{ and } q = \sqrt{\frac{(\gamma_0 - \gamma/2)^2}{4} - \Omega^2} \ . \tag{2}$$

In the above equation, $\gamma_0$ is defined as the natural linewidth ($1/T_1$) and $\gamma$ is the homogeneous linewidth ($2/T_2$). The orthogonal and parallel correlation functions in terms of $g^{(2)}(\tau)$ are given in[24] (see also supplementary information). Taking the analytical formula for $g^{(2)}_{\perp}(\tau)$ and $g^{(2)}_{\parallel}(\tau)$, the independently measured $T_1$ and $T_2$ times, and a convolution of the curves with our system response function, an excellent modelling (red curves) of all curves in Fig. 4 is obtained simultaneously. With this we obtain de-convoluted values of $g^{(2)}(0) = 0.08 \pm 0.02$, $g^{(2)}_{\perp}(0) = 0.53 \pm 0.03$, $g^{(2)}_{\parallel}(0) = 0.06 \pm 0.02$, and a visibility of $V_{HOM}(0) = 0.90 \pm 0.05$, respectively. This large visibility value of ~ 90 % demonstrates indistinguishability of photons within their coherence time and a nearly perfect overlap of the wave functions, showing the high potential of the source for quantum information science applications such as entanglement swapping[28].

We envision that the experiments can be extended to include a pulsed excitation scheme by using π-pulses to exactly excite one bright exciton state per pump cycle. In this scenario, the π-pulse spectral width has to be well matched to the linewidth of QD absorption in order to optimize the coherent excitation process and to minimize laser scattering. Prepared by such resonant triggered pumping conditions the QD will *deterministically* provide indistinguishable single photons, representing an essential ingredient for numerous future quantum information applications. With the presented QD-microcavity structures, Fourier transform-limited single-photon generation at repetition rates well inside the GHz regime is anticipated. With respect to the observed characteristic Mollow spectra in QD resonance fluorescence, another very appealing feature is the generation of background-free and temporally cascaded photon pairs between the two side wings of the 'dressed' emission triplet, if the laser is intentionally detuned from the QD resonance. As was first demonstrated on atomic fluorescence[29], this special emission scenario relies on a second-order scattering process, resulting in sequential single-photon emissions from the blue-shifted and red-shifted Mollow side peaks, respectively. This could serve as a two-color source of close to Fourier-transform-limited photons in computation schemes.

**Methods Summary**

We use single (In,Ga)As/GaAs quantum dots (QDs) coupled to the fundamental mode of surrounding all-epitaxial high-quality vertical cavity micro pillars for detailed optical studies in view of the generation process of indistinguishable photons at cryogenic temperatures (T ≥ 10 K). We investigate the fundamental conditions to create and verify 'ideal' photons at the ultimate Fourier transform-limit between coherence length $T_2$ and spectral bandwidth. For a direct comparison of spontaneous photon emission properties under incoherent versus fully resonant (i.e. coherent) excitation of individual QDs in micro pillars, a novel experimental geometry of orthogonal optical excitation and detection was developed. This technique provides an enhanced separation of resonance fluorescence from scattered laser signal with high signal-to-noise contrast. Using selective resonant s-shell excitation in conjunction with the tunability of the QD decay dynamics ($T_1$) via temperature-dependent emitter-mode coupling, the creation conditions of Fourier-limited photons with high $T_2/2T_1$ fidelity is investigated. The resonant character of QD fluorescence in terms of an observation of power-dependent Rabi frequency side bands in emission, i.e. the characteristic Mollow triplet, is traced in detail by the technique of scanning Fabry-Pérot high-resolution spectroscopy. Under resonant excitation into the s-shell, the indistinguishability of photon pairs is tested by polarization-resolved cw Hong-Ou-Mandel quantum interference measurements[30] on an asymmetric path-length, balanced Mach-Zehnder interferometer.

**Acknowledgements** We thank Alper Kiraz for fruitful discussions, Adriana Wolf and Monika Emmerling for their efforts to realize high quality cleaved sample structures, as well as Ata Ulhaq for supplementary measurements. We gratefully acknowledge financial support of the DFG via the research group "Quantum optics in semiconductor nanostructures".

**Figure 1: Experimental Setup. a,** Schematic representation of the experimental setups for combined low temperature (T > 9K) micro-photoluminescence (μ-PL), high-resolution photoluminescence (HRPL), and photon emission statistics measurements (second-order correlation $g^{(2)}(\tau)$ and two-photon interference (HOM) experiments). (BS: beamsplitter, PBS: polarizing beamsplitter, FPI: Fabry-Pérot interferometer, PH: pinhole, SM: polarization maintaining single-mode fiber, coll: collimator) **b,** Orthogonal geometry of excitation and detection on individual micro pillars as used in the experiments; Inset: side-view photograph of ordered micro pillars at the cleaved sample edge, in comparison with the (off-set) laser spot in the same plane.

**Figure 2: Characterization of single QD μ-PL. a,** Low-temperature (T = 10 K) photoluminescence spectra from the s-shell of a single QD close to the fundamental mode (FM) of a micro pillar cavity. The QD is quasi-resonantly excited via its p-shell. Temperature dependent QD-FM detuning is given by ΔE. **b,** High-resolution PL spectra (HRPL) of the same QD, scanned by a Fabry-Pérot interferometer with free spectral range FSR = 15 GHz ~ 62.035 μeV: The clear fine structure splitting ($\Delta E_{FS}$ = 11.0 + 0.2) μeV in emission results from an intrinsic anisotropy of this QD. **c,** Detuning-dependent luminescence decay time of the excitonic emission as measured by time-correlated single photon counting (TCSPC) under pulsed p-shell excitation conditions. At 24 K, zero QD-mode detuning is achieved, yielding a maximum $T_1$ reduction by factor 12 (Purcell). Also for non-zero detunings, a symmetric effect of lifetime reduction is traced. **d,** Frequency scanning signal obtained from the narrow band cw laser tuned over the s-shell of the single QD in Fig. 2a, which reveals two clear resonances with signal increase of up to factor 5 and fully resembles the emission fine structure $\Delta E_{FS}$ of Fig. 2b.

**Figure 3: Mollow triplets in resonance fluorescence of a single QD. a,** Power-dependent series of high-resolution micro-photoluminescence (HRPL) spectra observed from a single QD under cw resonant excitation into the s-shell plotted in log-scale. Whereas for weak s-shell excitation (bottom trace), the observed resonance spectrum consists of a single narrow emission peak with ~1.15 µeV FWHM homogeneous linewidth, Mollow triplets with increasing Rabi splitting energies from $\hbar\Omega$ ~ 6 µeV up to ~ 27 µeV can be clearly observed under increasing excitation power. **b,** the appearance of satellite peaks at frequencies $\omega_0$-$\Omega$ and $\omega_0$+$\Omega$ in the emission spectra under strong resonant excitation is a fingerprint of the formation of a "dressed" quadruplet of states, derived from the initial single ("bare") transition at $\omega_0$. **c,** Excitation power-dependent Rabi energies of the sidepeaks extracted from Fig. 3a. A perfect linear dependence on the square root of the excitation power is found as expected from theory.

**Figure 4: HBT and HOM measurements on the resonance fluorescence emission from a single QD. a,** Second-order correlation $g^{(2)}(\tau)$ measurement under resonant excitation of the QD at low intensity limit at 18 K. The de-convoluted value of $g^{(2)}(\tau)$ = 0.08 indicates a strong signal to background ratio (background ~ 4 %). **b,** and **c,** Two-photon interference measurements under orthogonal and parallel polarization settings of the interferometer arms, respectively. **d,** Two-photon interference visibility obtained from the data shown in **b** and **c**. Solid lines (red) in all figures are the corresponding correlation functions convolved with the measured system response of 400 $\pm$ 10 ps, and dotted lines are the de-convolved curves without considering any time limitation of the detectors. A large visibility value of $V_{HOM}(\tau = 0)$ ~ 0.90 is determined which indicates almost perfect mode overlapping and the high quality of the experimental conditions.

## Figure1:

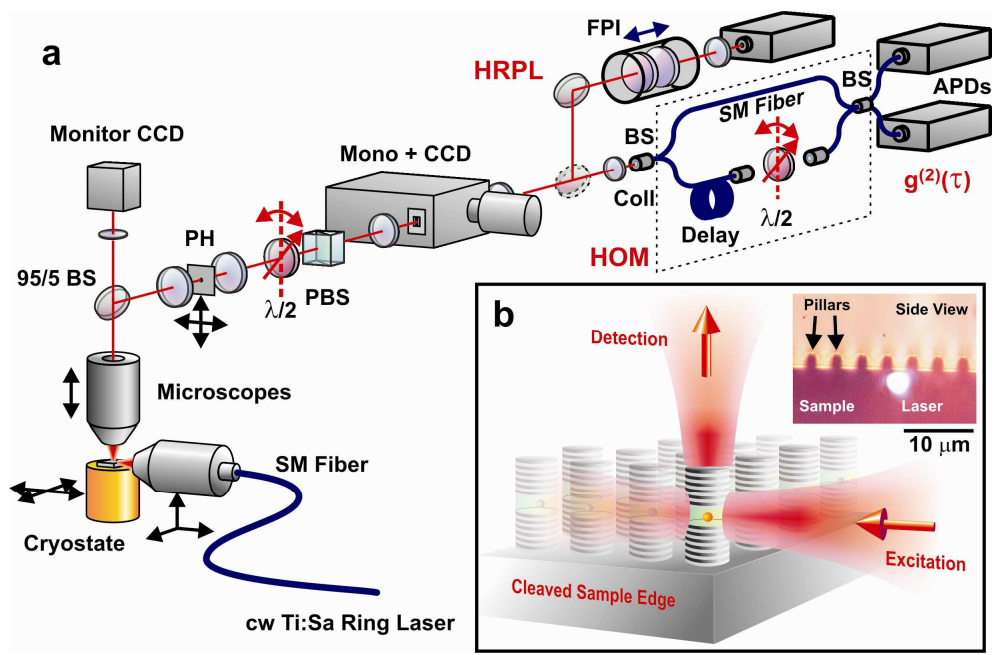

## Figure 2:

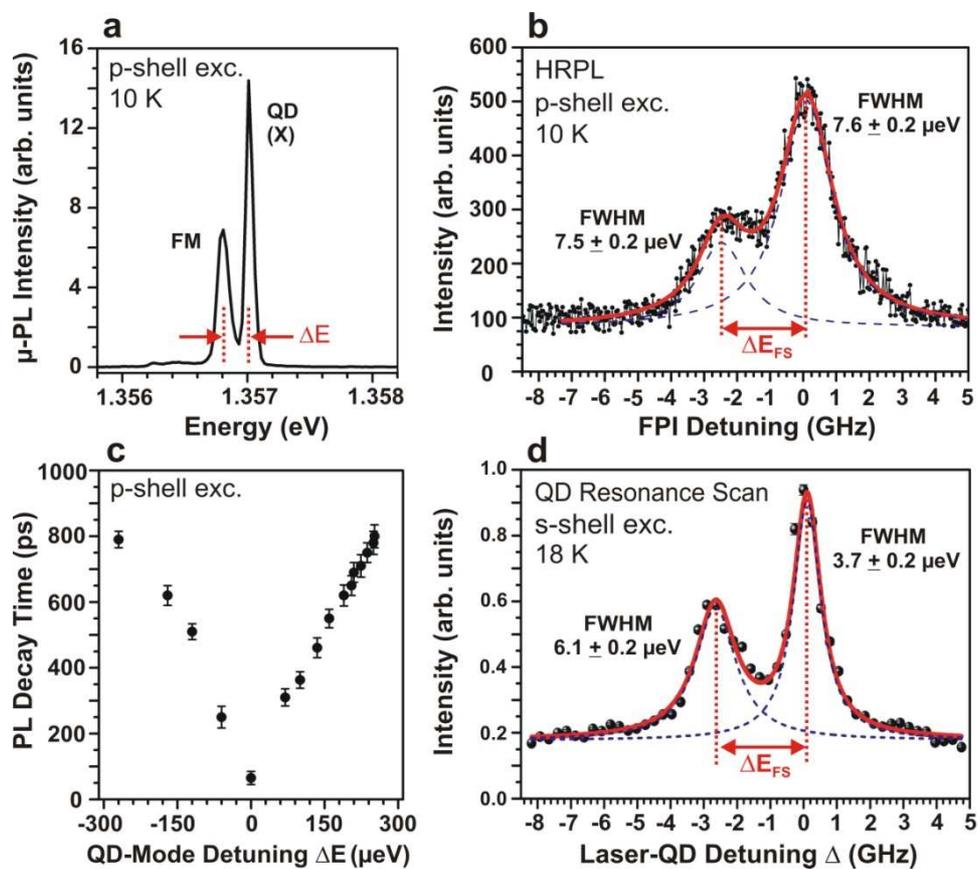

**Figure 3:**

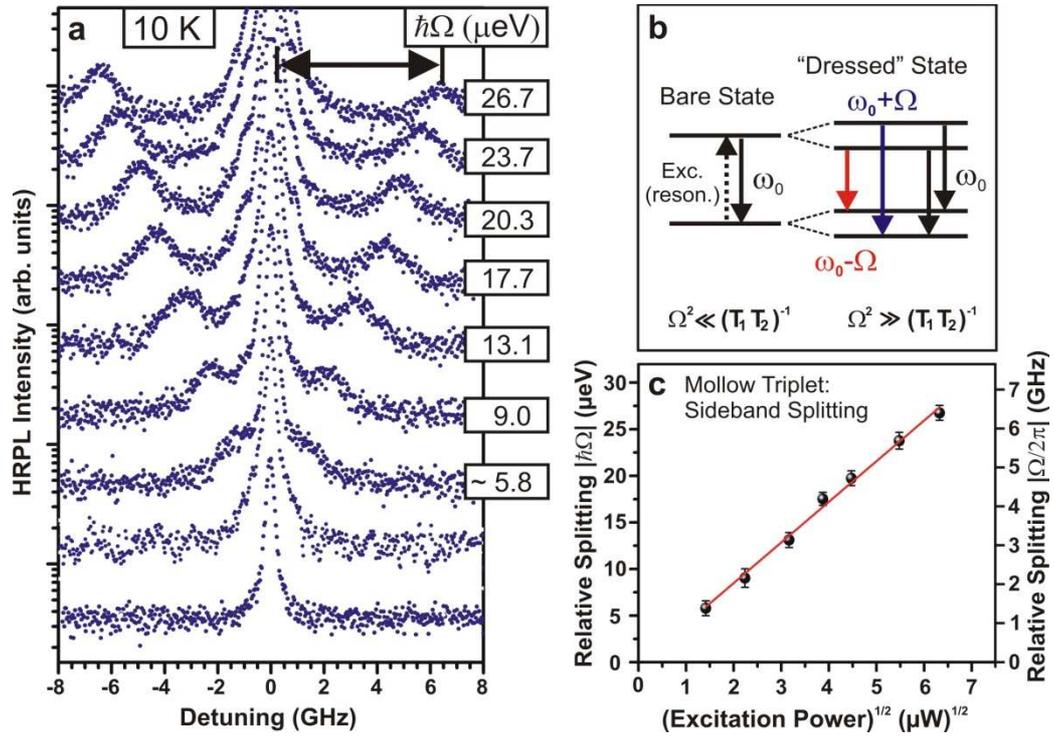

**Figure 4:**

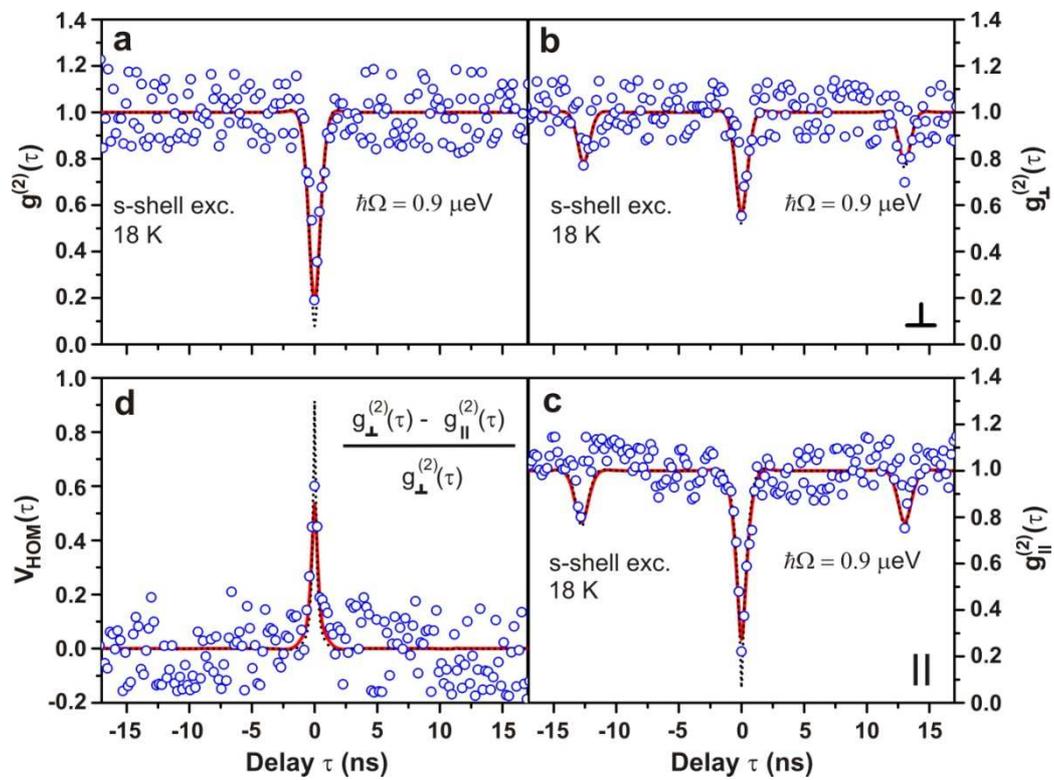

## Methods

**Sample Growth and Preparation**

Our samples were grown by molecular beam epitaxy on a GaAs substrate. The initial planar sample structure consists of a single layer of self-assembled (In,Ga)As QDs (In content of about 45 % and lateral density of ~$10^{10}$ cm$^{-2}$) which is positioned at the center of a 1λ-thick GaAs barrier layer. On top and below the central cavity layer, 26 and 30 periods of alternating λ/4-thick layers of AlAs/GaAs were grown to form distributed Bragg reflectors (DBRs), respectively. Micro pillar resonators were fabricated by a combination of electron-beam lithography and reactive ion etching. Details of the micro pillar processing can be found in[16]. A single pillar structure contains only 150-250 QDs on average which are additionally spectrally spread due to their inhomogeneous distribution within the whole dot ensemble (peak position at 1.37 eV; FWHM ~ 100 meV). For our studies we have cleaved the sample along the major orientation axes of the micro pillar arrays, leaving a lateral distance of less than 5 μm between the outer row of pillars and the sample edge. The sample was mounted in a cold-finger He-flow cryostat, allowing for variable temperature settings of T ≥ 10 K with ± 0.5 K stability.

**Experimental Setup**

Optical excitation of our micro pillar samples was provided by either a tunable narrow band (FWHM = 500 kHz) continuous-wave (cw) Ti:Sapphire ring laser or by a mode-locked tunable Ti:Sapphire pulse laser (76 MHz repetition of ~2 ps wide pulses) used in time-resolved micro-photoluminescence (μ-PL). As is schematically shown in Figs. 1a and b, a computer-controlled 3D-precision scanning stage equipped with a 50x microscope (fiber-coupled, SLWD, NA = 0.45) was horizontally adapted to our cryostat unit and used for selective optical excitation of QDs within their lateral growth plane. Together with an orthogonal detection geometry from top (i.e. along the vertical micro pillar axis), this setup allowed for a repetitive and long-term stable addressing of individual micro pillars under strong suppression of scattered laser stray light. The stray light suppression was additionally improved by use of adjustable pin-holes (PH) within the detection path towards the monochromator, which limit the area of detection in the focal plane to an effective spot diameter of ~2 µm. The collected light was spectrally filtered by a 1200 l/mm grating monochromator and send either to a scanning Fabry-Pérot interferometer (FPI; Finesse > 100) for high-resolution PL (HRPL) measurements with an enhanced spectral resolution of FWHM < 0.7 µeV, or to a fiber-based photon statistics setup. The latter assembly combines the techniques of two-photon interference and Hanbury Brown and Twiss-type (HBT) second-order auto-correlation measurements. Our photon statistics setup provides a time resolution of ~ 400 ps. Hong-Ou-Mandel-type (HOM) interference measurements on the indistinguishability of photons[30] are performed by an asymmetric Mach-Zehnder interferometer based on polarization-maintaining single-mode (SM) optical fibers. In this setup the collected and pre-filtered photon stream is first divided by a 50/50 non-polarizing beam splitter (BS) into two asymmetric paths, one of which imposes an extra fixed delay line of 13 ns (i.e., defining the interferometer asymmetry). Within the long optical fiber arm, an additional half-wave retarder plate is included to set the polarization of the photons mutually orthogonal or parallel with respect to the short fiber arm. The photons are finally interfered on a second 50/50 BS, the output ports of which are connected to two avalanche photo-diodes (APDs) for detection of photon coincidences. Without the Mach-Zehnder interferometer (highlighted by the dashed box in Fig. 1a), also $g^{(2)}(\tau)$ second-order photon auto-correlation measurements can be performed.

## Supplementary Information

### S1 Spectral properties under p-shell excitation

Detailed spectral properties of the non-resonantly coupled dot-cavity system have been investigated under p-shell (~ 900 nm) excitation conditions. All measurements in this work have been performed on a circular micro pillar structure with a diameter of 1.75 µm. The quality factor of the fundamental emission (FM) emission of the cavity (defined as the ratio of emission energy and the mode linewidth) has been experimentally verified from low excitation power spectrum as $Q = 13000 \pm 500$. Based on the measured Q-factor and the diameter of the micro pillar the theoretical value of the Purcell factor is estimated as $F_P \sim 50$. As is shown in Figs. 2a and S1 below, the QD emission is energetically on the higher energy side of the FM emission at 10 K with a detuning of about ~ 250 µeV. Increasing the temperature shifts the QD emission through the FM and a crossing behaviour is observed around the resonance (24 K) as a characteristic feature of the weak coupling regime (see Fig. S1).

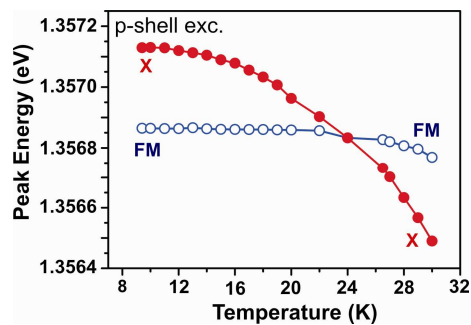

**Supplementary Fig. S1:** Temperature-dependent peak energies of the single exciton emission of the QD (X; filled circles) and the cavity mode emission (FM; open circles) under weak coupling, revealing a clear crossing behaviour at 24 K.

**S2 Spectral properties under s-shell excitation**

Spectral properties of the resonance fluorescence emission from the QD have been studied by using a high resolution PL (HRPL) setup (Fabry-Pérot Interferometer; see methods section). As is discussed in the main text, increasing the pump power changes the spectrum from a single emission line to Mollow triplets with a symmetrically increasing Rabi energy splitting $\hbar\Omega$ of the side peaks. The multi-Lorentzian spectrum at high powers is given by

$$F(\omega_{sc}) = \frac{3\gamma_{sp}/8\pi}{(w-w_{sp}-\Omega)^2+(3\gamma_{sp}/2)^2} + \frac{\gamma_{sp}/2\pi}{(w-w_{sp})^2+\gamma_{sp}^2} + \frac{3\gamma_{sp}/8\pi}{(w-w_{sp}+\Omega)^2+(3\gamma_{sp}/2)^2}$$

(S1)

with the radiative linewidths of $3\gamma_{sp}$ and $2\gamma_{sp}$ for the side peaks and for the central peak, respectively[S1].

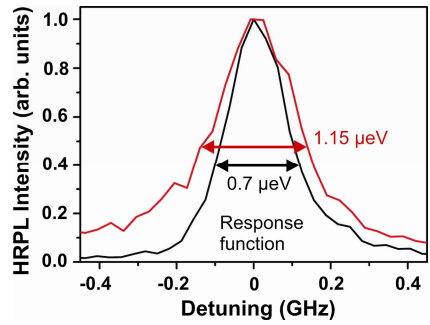

**Supplementary Fig. S2:** HRPL spectrum of the laser (910 nm) to illustrate the instrumental resolution of our FP interferometer, estimated as ~ 0.7 µeV (FWHM). The top trace (red line) represents the HRPL spectrum of the excitonic emission of the QD with a FWHM of 1.15 ± 0.05 µeV, obtained under low power resonant excitation.

The instrumental resolution of our full HRPL interferometer setup was derived as ~ 0.7 µeV (FWHM), according to a direct high resolution measurement on the narrow-band (500 kHz) cw laser at 910 nm (see supplementary Fig. S2). In comparison, we also plotted the HRPL spectrum of the QD resonance fluorescence emission at low pump power, which yields a FWHM of 1.15 ± 0.05 µeV, not being limited by the system resolution of our interferometer.

**S3 Two-photon interference measurements**

Two-photon interference measurements have been performed under cw resonant excitation conditions. Details of the fiber optics-based experimental setup are given in the methods section. As described before, one of the interferometer arms contains a half-wave plate to control the relative polarization of photons between the arms as orthogonal or parallel, thus making the paths distinguishable or indistinguishable. The second-order correlation functions for orthogonal and parallel polarization are given by[S2]

$$g^{(2)}{}_\perp(\tau) = 4(T_1^2 + R_1^2)R_2 T_2 g^{(2)}(\tau) + 4R_1 T_1 [T_2^2 g^{(2)}(\tau - \Delta\tau_2) + R_2^2 g^{(2)}(\tau + \Delta\tau_2)]$$

(S2)

and

$$g^{(2)}{}_\parallel(\tau) = 4(T_1^2 + R_1^2)R_2 T_2 g^{(2)}(\tau) + 4R_1 T_1 [T_2^2 g^{(2)}(\tau - \Delta\tau_2) + R_2^2 g^{(2)}(\tau + \Delta\tau_2)] * (1 - V e^{-\gamma|\tau|}),$$

(S3)

respectively. $R_1$, $R_2$, $T_1$, and $T_2$ are the reflectivity and transmission coefficients of the two fiber couplers, $\Delta\tau_2$ is the delay introduced in one of the arms, and V is a function of the overlap of the wave functions at the second beamsplitter. The visibility of the two-photon interference measurements is defined as:

$$V_{HOM}(\tau) = \frac{g^{(2)}{}_{\perp}(\tau) - g^{(2)}{}_{\parallel}(\tau)}{g^{(2)}{}_{\perp}(\tau)}. \tag{S4}$$

The instrumental response function (IRF) of our HBT setup has been measured as a Gaussian form with FWHM = 400 ± 10 ps. The above correlation functions (S2) and (S3) have been convoluted with this IRF and plotted as solid lines on Fig. 4.

In view of future applications, the extraction efficiency of indistinguishable photons from a cw resonantly excited micro pillar device is an important figure of merit. To give an estimation based on the experimental conditions of Fig. 4 (HOM), we refer to the theoretical maximum photon rate $f_{max}$ emitted by a single QD under a certain detuning from the micro pillar mode (introducing a specific Purcell enhancement), yielding $f_{max} = (T_1)^{-1}$ at saturation. From the total rate of detected photons, corrected for all specific efficiencies of the optical components in the setup, we can infer the actual rate of emitted (indistinguishable) photons f in relation to $f_{max}$. Especially for the experimental conditions of the data in Fig. 4, we derive a total photon extraction efficiency of the order of 1-2%. Worth to note, this efficiency might vary in dependence on the level of excitation (here: ~ 80%) with respect to saturation.

**S4 Future improvements for resonance fluorescence investigations**

As was discussed in the main text, the total emission signal under s-shell excitation is composed of resonance fluorescence from the QD and scattered laser light. The resonance fluorescence is principally limited by the effect of non-linear absorption saturation in a two-level system. From a perspective of future applications, the saturation level is the optimum efficiency regime of operation with respect to resonance fluorescence / indistinguishable photon generation. Under these conditions, effective techniques of scattered laser background light suppression are of highest significance.

As a possible to improve the signal-to-noise contrast for resonance fluorescence studies even at the level of emission saturation, we suggest to use a combination of alternative high-quality cavity geometries (e.g. rectangular-shaped micro 'mesas') and/or direct fiber coupling for excitation[S3] and orthogonal photon collection, which will increase the collection efficiency and serve as spatial filters with enhanced spatial selection and stray light suppression.